\documentclass[twocolumn,showpacs,amsmath,amssymb,showkeys]{revtex4}

\usepackage{graphicx}
\usepackage{dcolumn}
\usepackage{bm}
\usepackage{amssymb}

\begin{document}



\title{Delocalization induced by low-frequency driving in disordered
superlattices}

\author{Dario~F.~Martinez}
\author{Rafael~ A.~Molina}
\affiliation{Max-Planck-Institut f\"ur Physik komplexer Systeme,
N\"othnitzer Str. 38, 01187 Dresden, Germany}



\begin{abstract}
We study the localization properties of disordered semiconductor
superlattices driven by ac-fields. The localization length of the
electrons in the superlattice increases when the frequency of the
driving field is smaller than the miniband width. We show that
there is an optimal value of the amplitude of the driving field
for which the localization length of the system is maximal. This
maximum localization length increases with the inverse of the
driving frequency.

\end{abstract}



\maketitle


Real materials always contain a certain degree of disorder since
the atomic structure is never perfectly regular.
In fact, many physical properties are either influenced or even
mainly determined by this randomness. The understanding of the
effects of disorder in the physical properties of a material is
therefore of great practical importance and has played a central
role in condensed matter physics in the last half-century\cite{Lifshits}.

One of the most simple disordered systems is the motion of a
particle in a one-dimensional random potential. Realizations of
this system can be studied experimentally, e.g. in semiconductor
superlattices (SL) made by varying the alloy composition along one
of the dimensions of a compound semiconductor such as
Al$_X$Ga$_{1-X}$As \cite{Esaki}. With the technique of
molecular-beam epitaxy the degree of control of the disorder in
such systems can be very high. If the SL is long enough its
eigenstates are exponentially localized due to disorder. Electrons
in these localized states are spatially confined and their only
contribution to transport is through thermally activated hopping.
The main quantity of interest in this case is the localization
length $\lambda$ of the electron wave-functions, which is
controlled by the ratio between the bandwidth $\Delta$ and the
strength of the disorder $W$. A disordered system of length
$L>\lambda$ will behave as an insulator while a system with length
$L<\lambda$ will behave as a conductor\cite{Lifshits,Kramer}.

The fascinating effects of radiation on the transport properties
of SLs have been intensively studied both theoretically and
experimentally in the last three decades \cite{superlattices}. One
of the most interesting effects, first predicted by Dunlap and
Kenkre, is dynamic localization (as defined in \cite{Dunlap}), in
which the electron wave-function can be strongly localized in the
presence of an AC electric field. This effect was shown to be
associated with miniband collapse \cite{Holthaus92} which occurs
at the zeros of a Bessel function that depends on the field
amplitude. At these points the width of the miniband goes to zero,
the group velocity of a wave-packet vanishes and the electron
becomes effectively localized. Experimentally, this effect was
first observed as a suppression of current at some amplitudes of
the AC electric field \cite{Keay}.

When a time-periodic driving is applied to a disorder SL, a new
possibility emerges, i.e. controlling the localization of the
system by changing the amplitude and/or frequency of the driving
field. Holthaus \textit{et al}. \cite{Holthaus} showed that for
high frequencies, the Floquet states become localized (in contrast
with dynamical localization) and their localization (participation
ratio) depends on the ratio $\Delta_{eff}/W$, where $\Delta_{eff}
= \Delta J_0 (eVd/\hbar \omega)$. Here, $V$ and $\omega$ are the
amplitude and frequency of the driving field and $d$ is the
spatial period of the lattice.

In this letter we calculate, for the first time, the localization
length of a driven disordered SL and we focus on the previously
unexplored low frequency regime. For this purpose we use a
Floquet-Green function formalism which makes use of matrix
continued fractions\cite{Martinez}. This formalism allows us to
generalize the definition of $\lambda$ for time-periodic systems
and to calculate its value. We found that there are two distinct
and clear-cut regimes in this system: High-frequency regime for
$\hbar \omega> (\Delta+W)/2$ and low-frequency regime for $\hbar
\omega < (\Delta+W)/2$. In the first one, we found that $\lambda$
is a function of $\Delta_{eff}/W$, in perfect agreement with
previous works \cite{Holthaus}. In this regime, the driving field
always contributes to localize even more the wave-function (as
compared to the non-driven case). In contrast, in the low
frequency regime, we find that the driving can significantly
\textit{delocalize} the electrons. In addition to this new result,
we provide an intuitive explanation of why this should be so: For
low frequency, each additional Floquet channel created by the
driving provides the electron with new paths which will differ in
their degree of localization, with some having smaller and others
having greater localization length as compared to the non-driven
case. Since $\lambda$ is determined only by the path with the
greatest localization length, after the ensemble average has been
performed, additional propagation channels should always
contribute to increase $\lambda$. The same situation does not
occur in the high-frequency case because new paths introduced by
the absorbtion or emission of one or more photons, always have
localization lengths smaller than in the non-driven case.

We will model the one-dimensional disordered SL in the presence of
an AC field by a single-band Anderson Hamiltonian with diagonal
disorder \cite{Anderson} plus a time-periodic potential,
\begin{eqnarray}
\label{eq:Hamiltonian}
H= &-\frac{\Delta}{4}\sum_j \left(\left|j+1\right>\left<j\right|+
\left|j\right>\left<j+1\right|\right)+ \nonumber \\
& \sum_j \epsilon_j \left|j\right>
\left<j\right|+2V\cos{\omega t}\sum_j \left|j\right>j\left<j\right|.
\end{eqnarray}
The on-site energies $\epsilon_j$ are distributed uniformly from
$-W/2$ to $W/2$. The driving potential is due to an AC electric
field of amplitude $2V$. (The factor of 2 is for convenience.) In
a SL this term can represent THz radiation linearly polarized in
the growth direction of the lattice. The quantity $\Delta$ is
equal to the bandwidth of the non-driven system without disorder.
We will use units in which $\hbar=1$ and we set $\Delta=4$.

For a system that obeys discrete time-translational symmetry with
period $T$, there exists a
complete set of solutions to the Schr\"{o}dinger equation of the
form $\left|\Psi^{\alpha ,p}(t)\right>=\exp(-i e_{\alpha
,p}t/\hbar) \left|\phi^{\alpha,p}(t)\right>$, where
$\left|\phi^{\alpha ,p}(t)\right>=\left|\phi^{\alpha
,p}(t+T)\right>$. These are called the Floquet states of the
system and the periodic functions $\left|\phi^{\alpha
,p}(t)\right>$ obey an eigenvalue equation similar to the static
Schr\"odinger equation,
\begin{equation}
\label{eq:schrodinger-floquet} \left[H(t)-i\hbar\frac{d}{d
t}\right]\left|\phi^{\alpha,p}(t)\right>
=e_{\alpha,p}\left|\phi^{\alpha,p}(t)\right>,
\end{equation}
with $H(t)$ the time-periodic Hamiltonian of the system. The
eigenvalues of this equation can be written as $e_{\alpha,p}=
\epsilon_\alpha + p\hbar\omega$, for $0\leq\epsilon_\alpha \leq
\hbar\omega$ and $p$ an integer.
A Floquet-Green operator corresponding to Eq.
\ref{eq:schrodinger-floquet} can be defined and its Fourier
components satisfy
\begin{equation}
\label{eq:FloquetGreen} G^{(k)}(E)=\sum_{\alpha,p}
\frac{\left|\phi_{k+p}^{\epsilon_{\alpha}}\left>
\right<\phi_{p}^{\epsilon_{\alpha}}\right|}{E-\epsilon_{\alpha}-p\hbar\omega},
\end{equation}
where $\left|\phi_{p}^{\epsilon_{\alpha}}\right>$ are the Fourier
components of the Floquet eigenfunctions
$\left|\phi^{\epsilon_{\alpha}}\right>=\sum_p e^{-ip\omega
t}\left|\phi_{p}^{\epsilon_{\alpha}}\right>$. The transport
properties of driven systems have been formulated in terms of this
Floquet-Green operator which plays a similar role to the Green
operator in the Landauer formulism for conduction\cite{Kohler}.
We generalize the usual definition of the localization length in
terms of Green functions \cite{Kramer} and define the localization
length of the $kth$ Floquet component as
\begin{equation}
\label{eq:deflambdafloquet}
\frac{1}{\lambda^{(k)}(E)}=-\lim_{L\rightarrow \infty} \frac{1}{L}
\left<\ln\left|G^{(k)}_{1L}(E)\right|\right>.
\end{equation}
The different quantities $G^{(k)}_{1L}(E)$ are associated with the
probability of a process where an electron starts with an energy
$E$ at site $1$ and ends at site $L$ with energy $E+k\hbar\omega$.
Each one of these processes in principle can have a different
localization length associated to it. However, assuming that the
asymptotic behavior of the Floquet wave function is exponentially
decreasing, it is easy to see that the dominant term in the sum
over $p$ in Eq. (\ref{eq:FloquetGreen}) will always be the same,
independently of the value of $k$. Therefore the functions
$\lambda^{(k)}(E)$ are identical. From now on we will only compute
$\lambda(E)\equiv \lambda^{(0)}(E)$.

For the calculation of $G^{(0)}(E)$ we use a method developed by
one of the authors (see \cite{Martinez} for details). We want to
calculate the Floquet-Green operator for a periodic Hamiltonian of
the form $H(t)=H_0+2\cos(\omega t)V$, where $H_0$ and $V$ are any
time-independent operators in the Hilbert space of the system. The
Floquet components of the Green operator for this Hamiltonian
satisfy,
\begin{equation}
(E+k\hbar\omega-H_0)G^{(k)}-V(G^{(k+1)}+G^{(k-1)})=\delta_{k,0}.
\end{equation}
These equations can be solved using matrix continued fractions.
For the case $k=0$, one gets
\begin{equation}
G^{(0)}(E)=(E-H_0-V_{eff}(E))^{-1},
\end{equation}
where
\begin{equation}
V_{eff}=V^+_{eff}(E)+V^-_{eff}(E),
\end{equation}
with \begin{small}
\begin{equation} V_{{eff}} ^{\pm}(E)= V
\frac{1}{\displaystyle E \pm 1\hbar\omega-H_0 -V
\frac{1}{\displaystyle E\pm 2\hbar\omega-H_0
-V\frac{1}{~~\vdots~~}V}V}V~~. \label{eq:Veff}
\end{equation}
\end{small}

The convergence of equation (\ref{eq:Veff}) is system specific.
For our Hamiltonian, Eq.(1), the number of bands necessary to
ensure convergence increases linearly with $VL/\omega$. The
numerical performance of our method is determined by the speed in
the calculation of an $L \times L$-matrix inverse for each Floquet
sideband.
\begin{figure}
\begin{center}
\includegraphics[height=8cm,width=5cm,angle=-90]{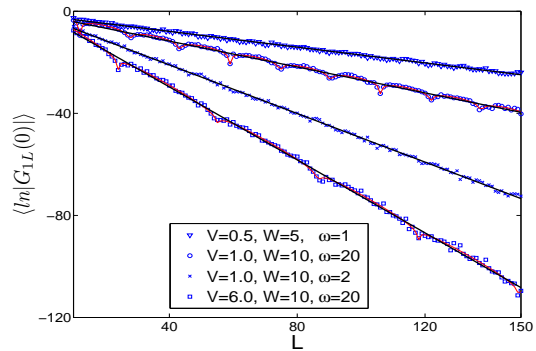}
\caption{\label{fig:scaling} (Color online) $\left<\ln
G_{1L}(0)\right>$ as a function of the length of the system $L$
for different values of disorder $W$ and frequency $\omega$.
High-frequency results, $\omega=20$, show regular oscillations on
top of the exponential decay, according to the equation
$G_{1L}(0)= AJ_{0}(2VL/\hbar\omega)\exp^{-L/\lambda}$. Low
frequency curves do not show these oscillations. The value of
$\lambda$ is the negative inverse slope of the linear fits.}
\end{center}
\end{figure}
For our disordered system we know that the localization length in
the middle of the band ($E=0$) for $V=0$ behaves as $\lambda_{0}
\approx 105\left(\frac{\Delta/4} {W}\right)^2$ \cite{kappus}.
In this case it is also known that $\lambda$ as a function of the
energy $E$ follows a parabolic law with a maximum at $E=0$. For
$|E|>(\Delta+W)/2$ the localization length decreases rapidly with
$|E|$. In this work, we will always take $E=0$.

In Fig.\ref{fig:scaling} we show some examples of the results
obtained for the ensemble average of $\ln ~G_{1L}(0)$ as a
function of the length of the system $L$. (We have dropped the
superscript in the Floquet-Green operator.) As expected, a
straight line fits the data very well. The wave-functions decay
exponentially with the distance. The negative slope of these
curves corresponds to the inverse of the localization length when
$L \gg \lambda$. It is important to point out that the deviations
from this exponential behavior are \textit{not} due to a deficient
ensemble average. For high-frequencies they show a regular
pattern, a manifestation of dynamic localization which shuts off
the tunnelling between the sites $L-1$ and $L$ at the zeros of
$J_0 (2VL/\omega)$. In fact, for high-frequency and $L\gg\lambda$
one can show that the Floquet-Green function can be expressed
analytically as $G_{1L}(0)= A J_{0} (2VL/\omega)\exp(-L/\lambda)$.
As shown in Fig. \ref{fig:scaling} for the second and last sets of
data (from top to bottom), the thin (red online) continuous line
representing this analytical result fits the numerical data very
well. For low-frequencies, the Green function also decays
exponentially. However, the deviations from this behavior cannot
be expressed analytically and seem to occur equally on both sides
of the straight-line fit.

\begin{figure}
\begin{center}
\includegraphics[width=8cm,height=5cm]{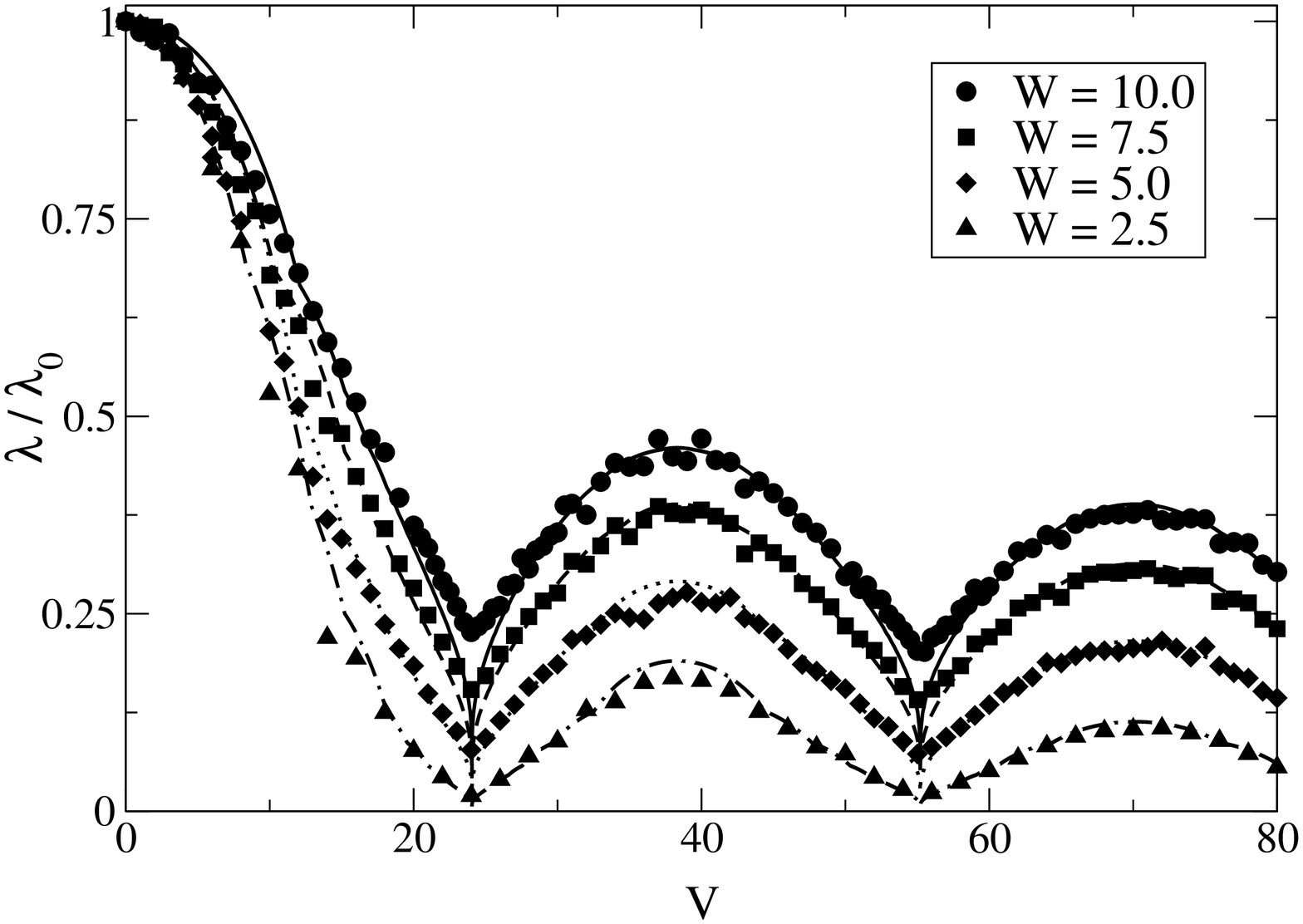}
\caption{ \label{fig:bessel} High frequency results. Results for
different values of disorder $W$ with $\omega=20.0$. The
corresponding results for a non-driven system with renormalized
bandwidth are shown with lines (obtained from numerical data).}
\end{center}
\end{figure}

We now show results for the high frequency regime, which has been
discussed in the literature \cite{Holthaus}, although to the best
of our knowledge, no calculation of $\lambda$ has been reported.
The high frequency limit can be characterized as the regime in
which the absorption or emission of any number of photons would
leave the particle with an energy outside of the region where the
eigenenergies of the non-driven system concentrate. This region is
well known to have a width $\Delta+ W$, and therefore, at the
center of the band this condition is satisfied when $\omega
> (\Delta + W)/2$. In the high-frequency
regime, the results for $\lambda$ are obtainable from the function
$\lambda_{0}(\Delta/W)$ but with a renormalized hopping term
$\Delta\rightarrow\Delta J_0(2V/\omega)$. Fig. \ref{fig:bessel}
shows, for $\omega=20.0$ and for several different values of
disorder, that the numerical data is in excellent agreement with
Holthaus' result. The minima of $\lambda$ correspond to the zeros
of the Bessel function $J_0(2V/\omega)$.

\begin{figure}
\begin{center}
\includegraphics[width=8cm,height=5cm]{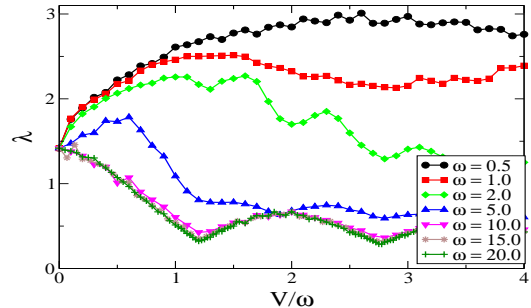}
\includegraphics[width=8cm,height=5cm]{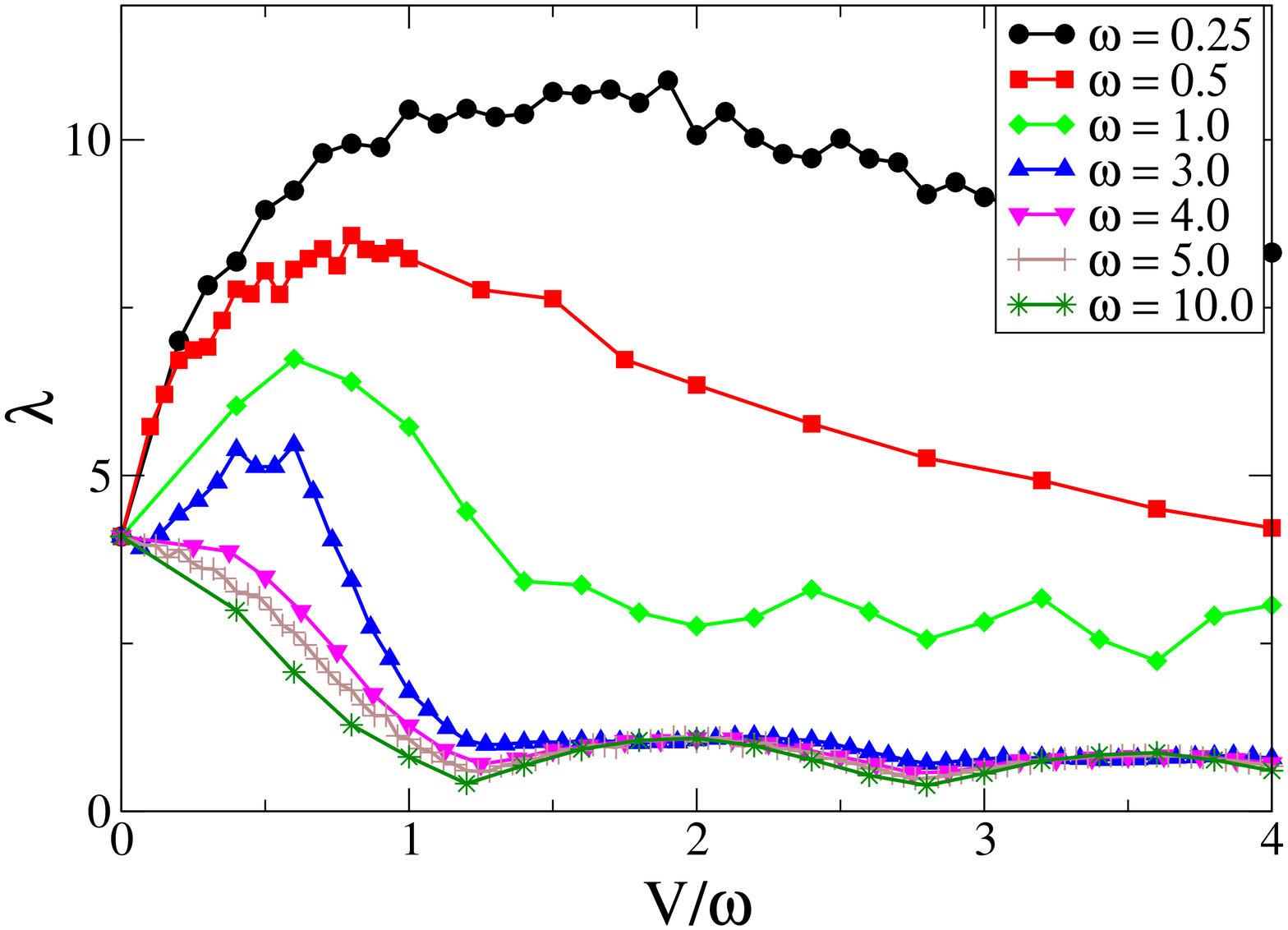}
\caption{\label{fig:hw} (Color online) $\lambda$ as a function of
$V/\omega$ for $W=10$ (top panel) and $W=5$ (bottom panel) and
different values of $\omega$.}
\end{center}
\end{figure}

In Fig.\ref{fig:hw} we show the localization length $\lambda$ as a
function of $V$ for low and high frequencies and for two values of
disorder, $W=10.0$ (top) and $W=5.0$ (bottom). Here, one can see
that there is a fundamental difference between those two regimes:
For $\omega<(\Delta+W)/2$ and $V$ small, $\lambda$ increases with
$V$, whereas for $\omega>(\Delta+W)/2$ it decreases. This
delocalization of the wave function due to low frequency driving
is the main result of this letter. For all the frequencies that we
explored in the low frequency regime, as $V$ was
increased $\lambda$ initially increased, then reached a maximum,
and finally decreased (with some oscillations). This behavior can
intuitively be understood if one assumes that the driving provides
the electron with new propagating channels $n$, each one with a
different localization length $\lambda_i$. According to this, one
expects $\lambda=max\{\lambda_1,\lambda_2,..\lambda_n\}$. Given
the linear dependence of the total number of Floquet sidebands
with $V$, it is expected that $n \propto V$, and therefore
$\lambda$ should initially increase with $V$. This is valid until
we reach the maximum number of effective channels that can be
supported in the region of the spectrum where the localization
lengths are comparable (size $\approx \Delta+W$). Here, $\lambda$
reaches a maximum value $\lambda_{max}$. Beyond this point,
$\lambda$ decreases with $V$ (with oscillations), as the weight of
the Floquet eigenstates oscillates and shifts to energies outside
the band.

\begin{figure}
\begin{center}
\includegraphics[width=8cm,height=5cm]{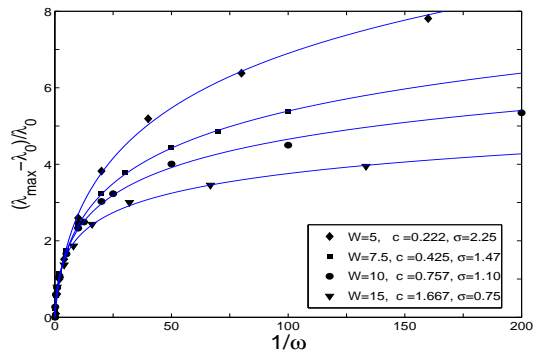}
\caption{\label{fig:eband} Behavior of $\lambda_{max}$ as a
function of $1/\omega$ for different values of disorder $W$. The
numerical data has been fitted using a function $\lambda
(n_{max},\sigma)$ derived from a simple statistical model where
the maximum number of effective channels is inversely proportional
to the frequency, $n_{max}=c/ \omega$ and $\sigma$ is the standard
deviation of each channel (with exponential probability
function).}
\end{center}
\end{figure}


In Fig. 4 we plot the data for
$(\lambda_{max}-\lambda_0)/\lambda_0$
vs. $1/\omega$. 
This figure suggests that low-frequency driving is more effective
in delocalizing a weakly disordered system than a strongly
disordered one. The continuous curves correspond to a fit to a
very simple mathematical model where the maximum number of
propagating channels is inversely proportional to the frequency
$n_{max}=c/\omega$. We also assume that each one these channels
has an exponential distribution function with standard deviation
$\sigma$ (a Gaussian distribution gives similar results). We can
see that this simple model fits the data very well. Despite of the
fact that some of the assumptions in this simple model are not
formally justified, we believe that its agreement with many of the
features in the numerical data support our intuitive explanation
for the delocalization of the wave-function at low frequencies. In
our system, $\lambda_{max}$ seems to increases monotonically with
the inverse of $\omega$. However, in a real SL a limiting
time-scale might appear, i.e. the energy relaxation time $\tau$.
If $\omega \tau < 1 $ the delocalization induced by the driving
field could be limited. For a typical relaxation time
$\tau=10^{-12}$ s and a typical miniband width $\Delta=3.5$ meV,
illumination with THz radiation of frequency $\omega = \Delta$
gives $\omega \tau \approx 5 > 1$.
This implies that a transition between the two frequency regimes
should be possible to implement with THz-irradiated SLs. For a
disordered SL of length $L \approx \lambda_0$ a change of several
orders of magnitude in the dynamic conductivity should be observed
when going from frequencies higher than the bandwidth to smaller
frequencies. Also, the behavior of the transport properties with
temperature should be very different, from a hopping regime with
some activation energy which is favored by higher temperatures to
a normal regime where transport processes should be hindered by
phonons and other temperature related effects. Other systems can
also be used to observe these delocalizing effects, e.g. atom
optical traps, which have been proposed as a testing ground of
Anderson localization. For this system, the lattice can be
implemented by far detuned counter-propagating laser beams, the AC
driving can be obtained using a periodic phase-shift between the
beams \cite{cold_atoms} and the random potential can be obtained
with a superimposed random speckle pattern \cite{speckles}.


In this work we have calculated the localization length for
one-dimensional SLs in a homogeneous AC electric field. We have
found two very different regimes according to whether the
frequency of the driving is smaller or bigger than the miniband
width of the disordered SL. For low frequency driving, the
localization length increases for moderate values of the driving
amplitude, while for high frequency the localization length
decreases due to the Bessel function renormalization of the
miniband width. We have shown that low-frequency driving can
increase the localization length by providing new propagation
channels. In this sense, the effect of such a field can be
compared to an increase in the dimensionality of the system, which
suggests that, in disordered systems where a metal-insulator
transition is expected when the dimension of the system increases,
low-frequency driving could have very dramatic delocalizing
effects.

The authors would like to thank H. Schomerus, G. Platero, R. A. Jalabert, and 
D. Weinmann for their suggestions regarding the presentation 
of this manuscript.

\end{document}